\documentclass[10 pt, conference]{IEEEtran}
\IEEEoverridecommandlockouts

\usepackage{fancyhdr}

\usepackage[noadjust]{cite}

\usepackage{amsmath}
\usepackage[linesnumbered,ruled,vlined]{algorithm2e}
\usepackage{algpseudocode}
\usepackage{graphicx}
\usepackage{tabularx}
\usepackage{textcomp}
\usepackage{xcolor}
\usepackage{balance}
\usepackage{comment}
\def\BibTeX{{\rm B\kern-.05em{\sc i\kern-.025em b}\kern-.08em
    T\kern-.1667em\lower.7ex\hbox{E}\kern-.125emX}}
\usepackage[english]{babel}
\usepackage[utf8]{inputenc}
\usepackage{subcaption}
\usepackage{caption}
\usepackage{multirow}
\usepackage{graphics}
\usepackage{adjustbox}
\usepackage{soul}
\usepackage{enumitem}
\usepackage{tabularx}
\usepackage{array}
\usepackage{wrapfig}
\usepackage{url}

\usepackage{booktabs}
\usepackage{graphicx}
\begin{document}


\title{BinSparX: Sparsified Binary Neural Networks for Reduced Hardware Non-Idealities in Xbar Arrays}



\author{\IEEEauthorblockN{Akul Malhotra and Sumeet Kumar Gupta}
\IEEEauthorblockA{
Purdue University\\
West Lafayette, Indiana, USA \\
malhot23@purdue.edu}}









\maketitle
\thispagestyle{empty}
\pagestyle{empty}

\begin{abstract}
  Compute-in-memory (CiM)-based binary neural network (CiM-BNN) accelerators marry the benefits of CiM and ultra-low precision quantization, making them highly suitable for edge computing. However, CiM-enabled crossbar (Xbar) arrays are plagued with hardware non-idealities like parasitic resistances and device non-linearities that impair inference accuracy, especially in scaled technologies. In this work, we first analyze the impact of Xbar non-idealities on the inference accuracy of various CiM-BNNs, establishing that the unique properties of CiM-BNNs make them more prone to hardware non-idealities compared to higher precision deep neural networks (DNNs). To address this issue, we propose BinSparX, a training-free technique that mitigates non-idealities in CiM-BNNs. BinSparX utilizes the distinct attributes of BNNs to reduce the average current generated during the CiM operations in Xbar arrays. This is achieved by statically and dynamically sparsifying the BNN weights and activations, respectively (which, in the context of BNNs, is defined as reducing the number of +1 weights and activations). This minimizes the IR drops across the parasitic resistances, drastically mitigating their impact on inference accuracy. To evaluate our technique, we conduct experiments on ResNet-18 and VGG-small CiM-BNNs designed at the 7nm technology node using 8T-SRAM and 1T-1ReRAM. Our results show that BinSparX is highly effective in alleviating the impact of non-idealities, recouping the inference accuracy to near-ideal (software) levels in some cases and providing accuracy boost of up to 77.25\%. These benefits are accompanied by energy reduction, albeit at the cost of mild latency/area increase.    
\end{abstract}

\begin{IEEEkeywords}

Binary neural networks, computing-in-memory, hardware non-idealities, IR drop, technology scaling, 

\end{IEEEkeywords}

\vspace{-0.1in}

\section{Introduction}
\label{sec:introduction}

Recently, there has been an immense interest in design techniques that can enhance the energy efficiency of deep neural networks (DNNs) and enable their deployment on the edge \cite{dl_edge}. From the algorithmic side, reducing the bit precision of DNN parameters via quantization lowers their energy, latency and storage demands, with minimal impact on accuracy \cite{quantization_zero}. Binary neural networks (BNNs) represent an extreme form of quantization wherein 1-bit weights and activations ($\in \{-1,+1\}$) are utilized, drastically improving the energy/area efficiencies \cite{bnn}.

From the hardware perspective, computing-in-memory (CiM), in which operations such as vector-matrix multiplications (VMMs) are performed within a crossbar (Xbar) memory array, is a promising direction \cite{cim}. CiM alleviates the massive data movement costs that plague standard von-Neumann-based DNN accelerators, leading to large energy and latency reduction. Another hardware aspect that is particularly important to meet the needs of growing DNN model sizes is technology scaling, which reduces the area and energy consumption, facilitating  further efficiency gains \cite{scaling1}. 

Utilizing binary quantization in conjunction with CiM combines the benefits of both the techniques and is highly suitable for edge computing. Several CiM-based BNN hardware designs (CiM-BNNs) have been developed utilizing CMOS as well as various non-volatile memory (NVM) technologies, showcasing substantial energy-latency-area benefits \cite{cim_bnn1, cim_bnn2, cim_bnn3}. 

Although CiM offers a significant energy/latency benefits, Xbar non-idealities due to wire resistance, driver/sink resistance and device non-linearities afflict the computational robustness \cite{pytorx,non_ideality2,non_ideality3}, leading to inaccurate VMM computations and degraded DNN accuracy. This issue is even more concerning in deeply scaled technologies, due to an increase in the wire resistivity and resistance \cite{interconnect1}.  To address the issue of Xbar non-idealities, various technological and design solutions have been proposed \cite{non_ideal_training1,non_ideal_training2,non_ideal_training3, compensate1, compensate2, write1,mapping1}. On the technology side, new interconnect materials/processes could potentially mitigate the Xbar non-idealities, but need more investigation \cite{interconnect1}. At the circuit/algorithmic levels, various techniques are being explored to improve CiM robustness, but most of them incur high training/finetuning costs \cite{non_ideal_training1,non_ideal_training2,non_ideal_training3} or a large performance penalty \cite{pwa}. More importantly, most of these works focus on higher-precision DNNs, leaving the analysis of Xbar non-idealities in CiM-BNNs largely unexplored. 

In this work, we establish that the effect of parasitic resistances on CiM robustness is severely exacerbated in BNNs (compared to the high precision DNNs) due to their unique properties. Thus, to harness the benefits of CiM-BNNs, especially in scaled technologies, there is a pressing need to develop techniques that can pointedly mitigate this issue.  

To that end, we propose BinSparX, a training-free technique that alleviates the impact of Xbar non-idealities on CiM-BNN accuracy. BinSparX reduces the magnitude of the CiM outputs of Xbar arrays by statically and dynamically sparsifying the BNN weights and activations, respectively. In the context of BNNs, we define sparsification as reducing the number of +1 weights and activations (or increasing the number of -1 values - mapped to 0 in hardware \cite{nand_net}). Hence, the Xbar output current is lowered, leading to reduced IR drops in the parasitic resistances. The key contributions of our work include:

\begin{itemize}
    
    \item We show that the unique attributes of CiM-BNNs make them highly prone to Xbar non-idealities, resulting in huge accuracy loss, especially in scaled technologies.  
    
    \item We propose BinSparX, a training-free non-ideality-mitigating technique tailored for CiM-BNNs.   

    \item  We demonstrate the effectiveness of BinSparX by applying it on 8T-SRAM and 1T-1ReRAM (resistive RAM) based CiM-BNNs designed in the 7nm technology node. 

    \item We evaluate the hardware implications of BinSparX showing energy benefits at mild latency/area overheads. 
    
\end{itemize}

\section{Background and Related Works}
\label{sec:background}

\subsection{CiM-based BNN hardware (CiM-BNNs)}
\label{sec:cim_bnn}

CiM-BNNs derive their high energy efficiency by (i) restricting the weights and activations to 1 bit each with values of +1 or -1 and (ii) performing in-memory VMM. The signed binary representation maps the scalar multiplication between weights and activations to an XNOR operation. To enable XNOR-based CiM, previous works have introduced customized bitcells \cite{cim_bnn1, cim_bnn2}, albeit at significant area and energy costs compared to standard bitcells. To avert these costs, recent works, such as NAND-Net \cite{nand_net}, have shown that by applying linear transformations to activations and weights, the XNOR operation can be converted into an AND operation, which can be implemented using standard memory bitcells. These transformations map the original weights ($W$) and activations ($I$) to $W'$ and $I'$, respectively, translating from \{'-1','1'\} to \{'0','1'\} domain. These transformations are given by: $I = 2I' - 1$  and $W = 2W' - 1$. Hence, the new weights and activations are represented with high/low resistance state (HRS/LRS) of the memory and binary voltages, respectively, leading to seamless AND-CiM with standard memories. Using these transformations, the dot product can be written as:
\begin{align}
    &\sum_{i=1}^{n} I_i W_i = 4\sum_{i=1}^{n} I_i'W_i' - 2 \sum_{i=1}^{n} I_i' - 2 \sum_{i=1}^{n} W_i' + n     
    \label{eq:nand_net}
\end{align}
Here, $n$ is the length of the $W$ and $I$ vectors. While $\sum_{i=1}^{n} I_i'W_i'$ can be efficiently computed with AND-CiM macros, some pre- and post-processing is needed to obtain $\sum_{i=1}^{n} I_iW_i$. First, a near-memory adder tree is needed to compute $\sum_{i=1}^{n} I_i'$. Note that since $\sum_{i=1}^{n} W_i'$ and $n$ in (1) are fixed, they can be pre-computed and stored.  Also, since an activation vector is shared by many weight vectors spread across multiple Xbar arrays, the cost of the adder tree is amortized \cite{nand_net}. Second, some arithmetic circuitry is needed to add/subtract the second, third and fourth terms in (1) from $\sum_{i=1}^{n} I_i'W_i'$ (CiM output). These additional circuitry generally incur much lower cost compared to custmized bitcells.  Given the multiple benefits of NAND-Net over customized bit-cells, in this work, we focus on NAND-Net based CiM-BNNs.

\subsection{Techniques to mitigate Xbar non-idealities}
\label{sec:non_idealities}
While Xbar arrays seamlessly faciliate CiM of VMM, they suffer from various non-idealities such as IR drops in the parasitic resistances, device non-linearities etc., which lead to CiM errors, impairing the inference accuracy. 

Several previous works have explored techniques to alleviate the impact of hardware non-idealities on CiM-based DNNs. On the algorithmic side, \textit{non-ideality aware} training, fine-tuning and statistical correction has been explored \cite{non_ideal_training1,non_ideal_training2,non_ideal_training3, compensate1, compensate2}. However, these solutions require access to labelled training data/data statistics and large computational resources, which may not always be available.

From the hardware perspective, novel write schemes, redundancy, non-ideality-aware weight mapping and partial wordline activation (PWA) are commonly used non-ideality-mitigating strategies \cite{write1,compensate2,mapping1}. For example, \cite{redundant1} and \cite{compensate2} use a redundant row per array to enhance CiM robustness, albeit at the cost of area  \cite{compensate2}. Works based on non-ideality-aware mapping \cite{mapping1,wagonn} reduce IR drops via row/column re-arrangement but have applicability limited to only certain Xbar designs or memory technologies. PWA reduces the current in Xbar arrays, but at the cost of latency \cite{pwa}.

Furthermore, two important points should be noted. First, \textit{most} of the aforementioned works have been evaluated for pre-45nm technology nodes and their effectiveness in deeply scaled technologies requires further study. Second, these works focus primarily on high-precision DNNs, leaving the impact of Xbar non-idealities in BNNs and their mitigation unaddressed. In this work, we fill this gap by evaluating the Xbar non-idealities in CiM-BNNs at the 7nm technology node and proposing BinSparX to enhance CiM-BNN accuracy.    
\section{BNN Sparsity and Xbar Non-Idealities}
\label{sec:observations}

In this section, we describe the issue of aggravated Xbar non-idealities in CiM-BNNs that stems from the distinct attributes of BNNs. For this, we take ResNet-18 BNN as a case study, and analyze the Xbar outputs, current deviations due to Xbar non-idealities and their impact on BNN accuracy.  

\subsection{Partial sum analysis for CiM-BNNs}
\label{sec:partial_sum_analysis}

We start by profiling the expected CiM output of the Xbar arrays ($\sum_{i=1}^{n} I*W$), which we refer to as the \textit{ideal} partial sums.
Note, DNN weight matrices are typically partitioned and mapped onto multiple Xbars. The partial-sums produced by these Xbars are combined to obtain the VMM output.

For our partial sum analysis, we map a ResNet-18 BNN trained on CIFAR-10 on Nand-Net-based 64x64 size Xbar arrays. For comparison, we also map a ResNet-18 4-bit DNN on 64x64 Xbar arrays. For this, we store the 4-bit weights in their 2's-complement form with each of their bits stored in a binary memory element (to be compatible with designs such as SRAMs). The 4-bit activations (which are non-negative due to ReLU activation in high precision DNNs) are bit-streamed using binary voltages (0 and $V_{DD}$). We profile the distribution of the \textit{ideal} partial sums (values ranging between 0 and 64) by performing inference using the entire CIFAR-10 testset.  

\begin{figure}[t!]
\centering
  \includegraphics[width = 0.95\linewidth]{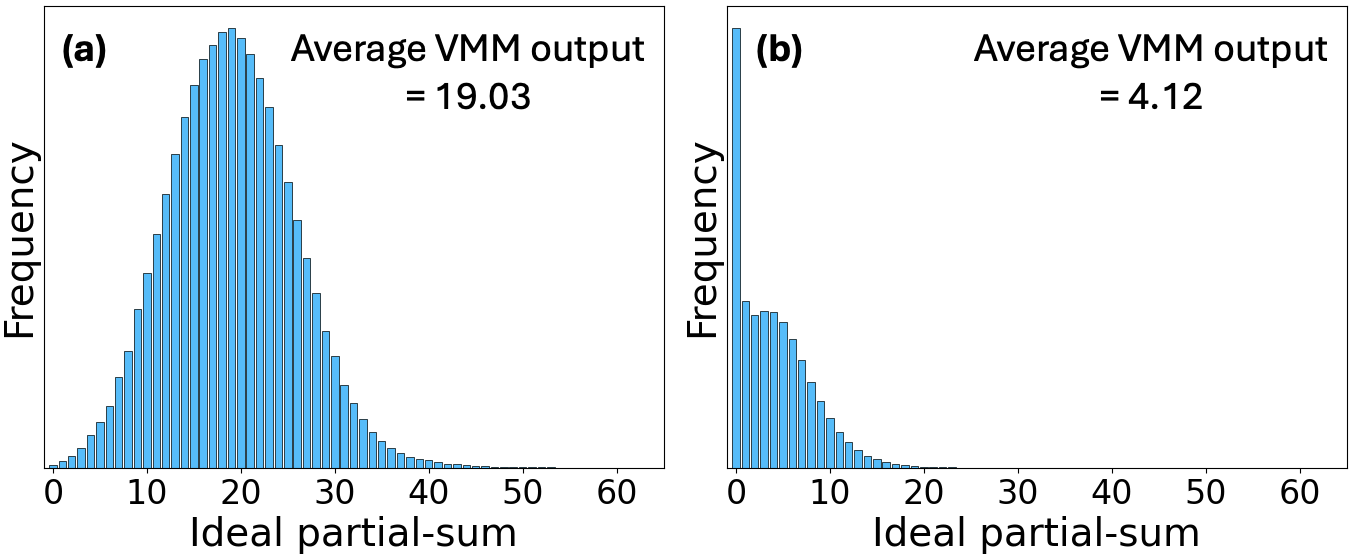}
  \caption{Histograms showing the frequency of ideal partial-sums in a (a) ResNet-18 BNN and (b) ResNet-18 with 4-bit weights and activations. The Xbar array size is 64x64. We observe that BNN produces larger partial-sums than the 4-bit DNN. }
  \label{fig:histograms}
\end{figure}

  Our analysis (Fig.~\ref{fig:histograms}) shows that BNNs produce substantially larger partial-sums than the 4-bit DNNs. The \textit{average} BNN partial-sum is $\sim$ 19, whereas that for the 4-bit DNN is $\sim$ 4. We attribute this to two distinct properties of BNNs. First, due to \textit{only two possible values} (+1 and -1) of weights and activations, the distribution of 1's and -1's (0's in Xbars) in BNNs is more or less uniform. This is also true for other BNNs used in our work (Fig.~\ref{fig:binsparx_reduction}). In contrast, the higher precision DNNs have a zero state and exhibit an approximately normal distribution of DNN weights (and in some cases, activations) centered around that zero state. Additionally, quantization, even to standard 4-8 bits, collapses close-to-0 values to 0. As a result, high precision DNNs exhibit larger input/weight sparsity compared to BNNs. Second, bit-streaming of activations coupled with the ReLU activation function (both of which are absent in BNNs) increases the sparsity in higher precision convolutional DNNs. These two distinctions between BNNs and higher precision DNNs culminate in substantially higher partial-sums in CiM-BNNs. It is important to note that while the second difference may not hold for non-convolutional DNNs, such as transformers, the first distinction remains valid.     

The inherently high partial sums in BNNs imply large currents in the Xbar columns, which are expected to severely worsen the hardware non-idealities (such as IR drops). To understand this, we evaluate the Xbar non-idealities and their impact on inference accuracy of ResNet-18 BNN. Before we discuss this, let us describe our evaluation framework.

\subsection{Simulation framework for evaluating Xbar non-idealities}

\label{sec:non_idealities_framework}

\begin{figure}[t!]
\centering
  \includegraphics[width = 0.95\linewidth]{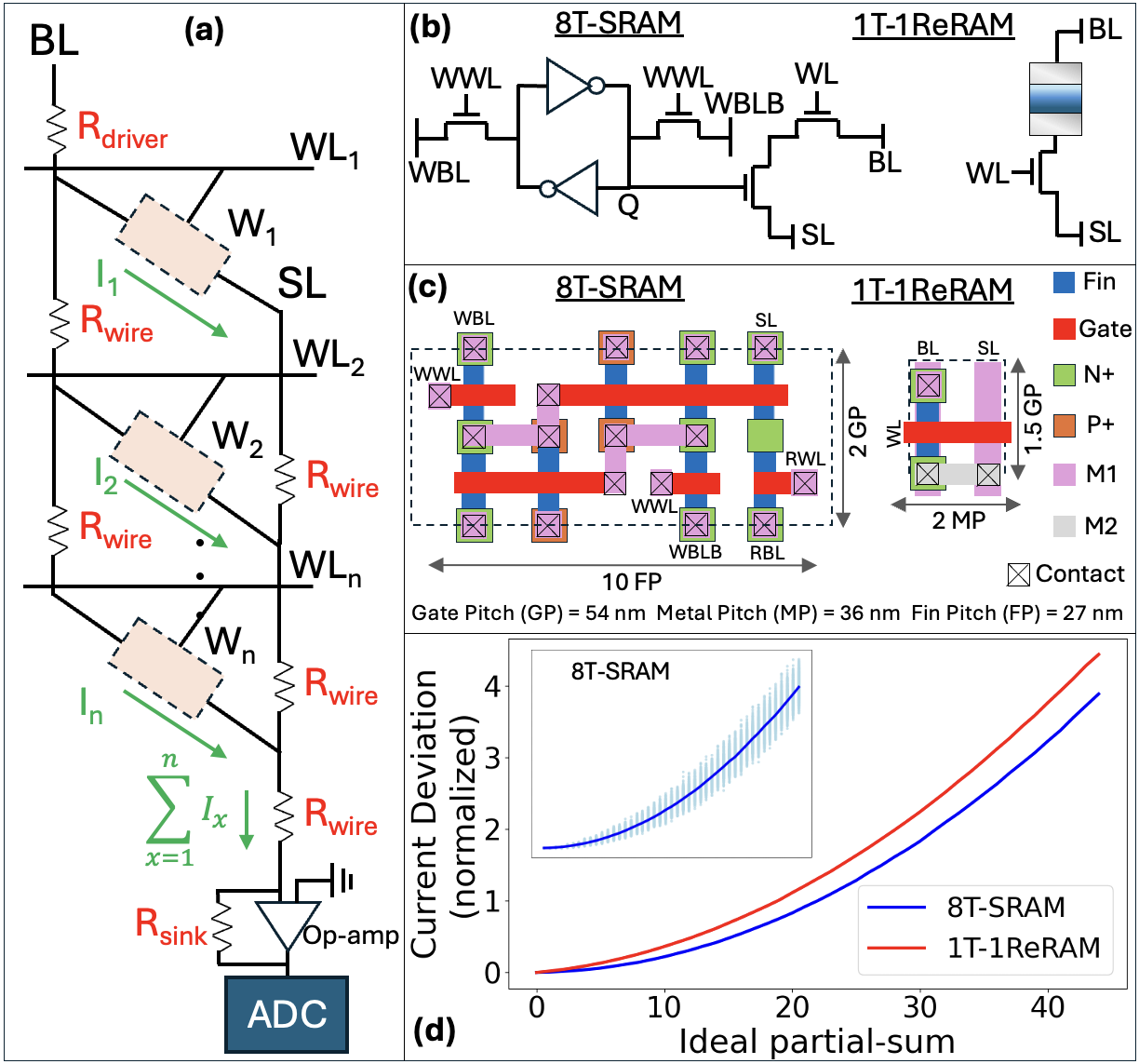}
  \caption{(a) Xbar array column with non-idealities. (b-c) schematics and layout of 8T-SRAM and 1T-1ReRAM bitcells. (d) average current deviation (normalized to current quantum between adjacent ADC levels) versus partial-sum. The inset shows the range of current deviations for each partial sum corresponding to different input-weight combinations. Current deviation increases superlinearly with rising partial-sum.}
  \label{fig:non_idealities}
\end{figure}

Our simulation framework is based on a customized Xbar simulator, which exactly and self-consistently solves the Kirchoff's current/voltage laws in conjunction with the memory device models to produce the output currents of the Xbar. Fig.~\ref{fig:non_idealities}(a) shows a column of the Xbar array, which our simulator models.  Our framework captures non-idealities such as wire resistance ($R_{wire}$), driver resistance ($R_{driver}$), sink resistance ($R_{sink}$) and device non-linearities. In this work, we utilize the design where the activations ($I$) are applied on the gates of the access transistors (as it has been shown to be more resilient to hardware non-idealities than other design options \cite{neurosim}). As a result, there is negligible steady-state IR drop along the row. Hence, we can analyze the impact of non-idealities on each Xbar column independently \cite{non_idealities_chunguang}. Note that we use an op-amp to sense the output current before feeding it into an analog-to-digital converter (ADC), mitigating the effects of sink resistance, as proposed in \cite{multibit}. Such choices enable us to analyze baseline BNNs designed for improved CiM robustness. We show that despite these choices, the inherently high partial sum leads to unacceptably low accuracy in BNNs.  

In this work, we analyze Xbars using two memory technologies:  8T-SRAM and 1T-1ReRAM ( Fig.~\ref{fig:non_idealities}(b) and (c)). We use 7nm predictive models \cite{ptm} for the transistors and the experimentally calibrated model in \cite{rram} for ReRAMs. For wire resistivity (both line metals and vias), we use models (and their validated parameters)  from \cite{xinkang}, which capture surface scattering, grain-boundary scattering, and the effect of liner/barrier layers. For an optimistic baseline, we choose interconnects with scaled liners (lower resistance than standard). The wire resistance per bitcell is calculated by multiplying the  bitcell height (Fig.~\ref{fig:non_idealities} (c)) with the resistance per unit length of the bitline (BL) and sense-line (SL). 

By virtue of the models discussed above, our framework seamlessly captures the non-idealities due to finite current $I_{HRS}$ and $I_{OFF}$ produced by bitcells when $I = 1$ and $W = 0$ and $I = 0$ and $W = 0/1$, respectively. Note that for 8T-SRAM, the ratio of $I_{ON}$ (current produced when $I = W = 1$) to $I_{OFF/HRS}$ is large ($> 10^4$). However, for 1T-1ReRAM, $\frac{I_{ON}}{I_{HRS}}$ is in the range of 10-50 \cite{non_idealities_chunguang}, leading to CiM errors under certain scenarios. To mitigate the impact of $I_{HRS}$ in 1T-1ReRAM Xbar, we use a dummy column, as in \cite{dummy_column, non_idealities_chunguang}. 

We obtain the individual bitcell current as a function of its terminal (WL, BL and SL)  voltages from bitcell simulations in SPICE and form look-up tables (LUTs). The LUTs are used in conjunction with the customized Xbar simulator described above to obtain the non-ideal output current. We implement our simulator in Pytorch, facilitating its seamless and efficient integration into DNN workloads. The customized simulator has been extensively validated with SPICE simulations of Xbar arrays, showing a \textit{maximum} error of 0.3\%.

\subsection{Impact of large partial-sums on Xbar non-idealities}

Utilizing this simulation framework,  we first analyze the current deviations as a function of expected partial sum output (Fig.~\ref{fig:non_idealities}(d)). In this section, we choose the design points (such as routing metal layer M6 for BL/SL, optimized ON currents etc. - more details later) that offer the highest accuracy for the BNN (optimistic baseline). In Section V, we will carry out a more comprehensive discussion for different design points.  For \textit{each} ideal partial-sum $x$, we obtain the non-ideal currents from our simulator for 10,000 unique weight-activation pairs.  It can be seen in Fig.~\ref{fig:non_idealities}(d) that the \textit{average} current deviation for both memory technologies not only increases with an increase in $x$, but grows at a \textit{superlinear} rate. As $x$ increases, there are more 'ON' bitcells in a column, leading to higher BL/SL current and larger IR drops.   

Since inherently, BNNs exhibit higher partial sums (average ideal value $\sim$ 19) than higher precision DNNs, the current deviations (due to the superlinear trend) are much more severe, leading to large number of CiM errors. We show its impact on CiM-BNN accuracy by evaluating ResNet-18 BNN with non-ideal Xbars on CIFAR-10. Our results show that the accuracy for 8T-SRAM and 1T-1ReRAM drops from 88.50\% (software accuracy) to 52.09\% and 18.08\%, respectively. As discussed later, other design configurations yield even lower accuracy values. These drastic accuracy drops, due to the large partial sums in BNNs, deem the CiM-BNN unusable, especially in deeply scaled technologies. In the next section, we propose BinSparX, a training-free technique that  mitigates this issue.


\section{BinSparX: The proposed solution}
\label{sec:method}

BinSparX reduces the partial sums in BNN Xbar arrays by employing two operations, namely static weight sparsification and dynamic activation sparsification. Both these operations capitalize on the \textit{binary-valued} weights and activations in BNNs that facilitate some weight/activation-sparsifying transformations while ensuring no effect on the VMM functionality.   

\subsection{Static weight and dynamic activation sparsification}

Recall, in the NAND-Net architecture (described in Section~\ref{sec:cim_bnn}), -1 and +1 weights are mapped to 0 (HRS) and 1 (LRS), respectively in the Xbar arrays. Similarly, -1 and +1 activations are mapped to 0 and $V_{DD}$, respectively. Therefore, if the number of -1's in the weight matrix or activation vectors are increased, it would imply increasing the zero weight bits stored in the Xbar or the zero input activation voltages applied to the wordlines (WL). In this section, we define weight and activation sparsification in this context i.e. reducing the number of +1's (or increasing the number of -1's) so that the Xbar array or the WL voltage vectors are sparsified. Such a weight/activation sparsification would naturally reduce the partial sums and reduce the impact of non-idealities.     

\begin{figure}[t!]
\centering
  \includegraphics[width = 0.92\linewidth]{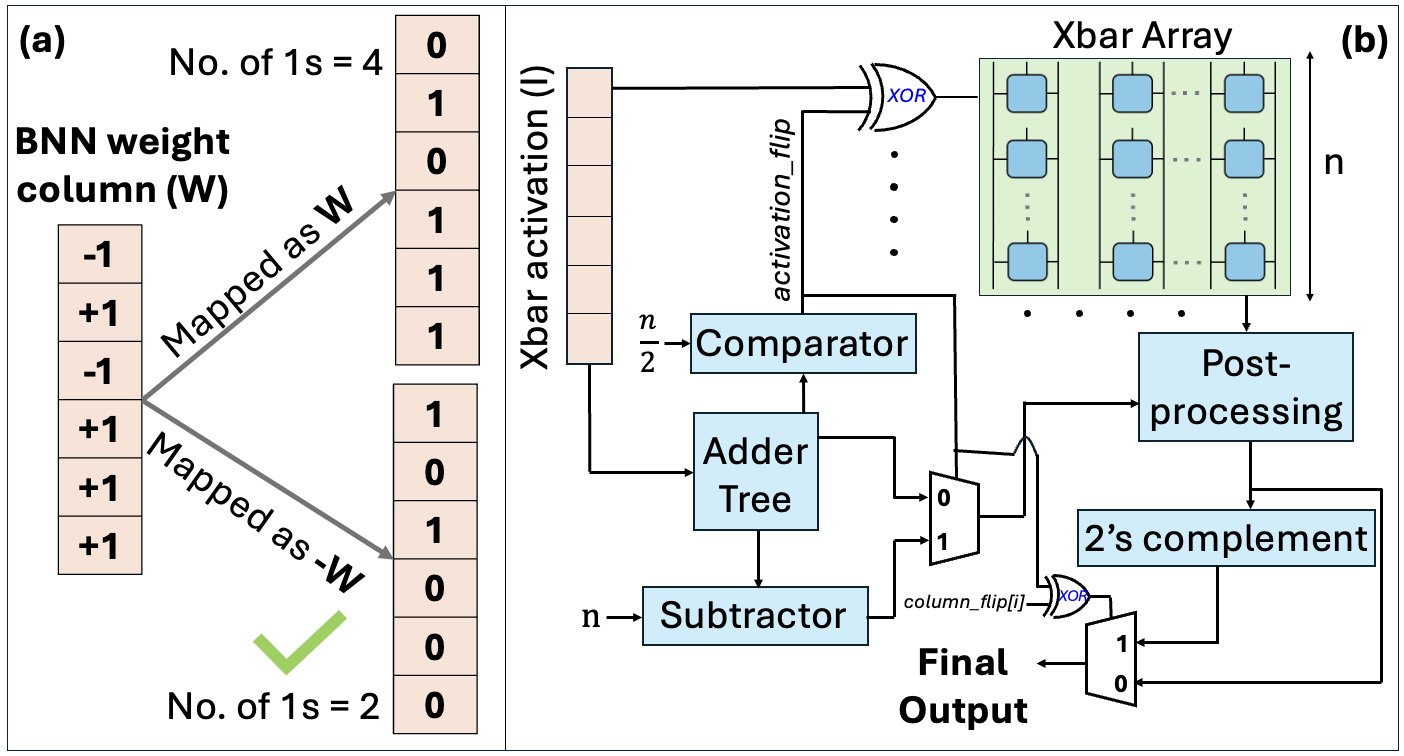}
  \caption{(a) Static weight sparsification being applied on a BNN weight column. (b) Hardware implementation of BinSparX.}
  \label{fig:binsparx}
\end{figure}

Fig~\ref{fig:binsparx}(a) illustrates the proposed static weight sparsification operation utilized by BinSparX for an Xbar column.  The key role of this operation is to minimize the number of 1s in each column of the weight sub-matrix ($W$) \textit{before} they are mapped onto an Xbar column. We achieve this by first calculating $\sum W$ and checking whether it is $\geq$ 0. If yes, then there are more +1s than -1s in $W$. In that case, we multiply all elements in $W$ by -1 (flip). If $\sum W < 0$, we keep $W$ unchanged. We also keep track of whether or not a flip has been performed in a column using a $column\_flip$ vector. If the $i^{th}$ weight column is stored as $-W$ ($W$), $column\_flip[i]$ is equal to 1 (0). Once applied, the modified weights are deployed on the Xbar arrays and the $column\_flip$ vector is stored in a peripheral register. For an Xbar with $n$ columns, $column\_flip$ is of size $n$ bits.  The $column\_flip$ vector is used to maintain the correct VMM functionality. If $column\_flip[i]$ is equal to 1,  the corresponding VMM output of the column is multiplied by -1 (since $\sum I*W = -1\cdot \sum I*(-W)$). Otherwise, the VMM output is used as it is.

Thus, by choosing between storing $W$ or $-W$ based on $\sum W$, we enhance the sparsity in the Xbar weights. Note that we refer to the weight sparsification operation as \textit{static}. This is because the BNN weights are constant during inference, enabling us to perform this step in advance (i.e. in software) rather than during inference.

Similar to the weight sparsification, activation sparsification is performed by selecting between activation sub-vectors $I$ or $-I$ based on $\sum I$. Here, $I$ is the sub-vector that is applied as an input to an Xbar array and has the size equal to the number of rows of the Xbar array ($n$). It is important to note that unlike weights, activations are dynamic and vary with each input, requiring sparsification \textit{during} inference. Thus, the selection between $I$ or $-I$ must be performed in the hardware before the modified activation vectors can be applied as WL voltages. Recall, the NAND-Net hardware utilizes $I$ in the [0,1] domain. Therefore, we compare $\sum I$ to $\frac{n}{2}$. If $\sum I > \frac{n}{2} $, there are more 1s than 0s (equivalent to -1s) in $I$, and we flip every element of $I$. Otherwise, $I$ is unchanged. We name the output signal of the comparator  $activation\_flip$. If an activation sub-vector is flipped, $activation\_flip$ is 1; else it is 0. $activation\_flip$ is utilized to ensure the correct VMM functionality. If $activation\_flip$ is 1, we negate the partial sums obtained from all the columns of the corresponding Xbar array (i.e. on which the flipped input is applied). This is because $\sum I*W = -1\cdot \sum (-I)*(W)$. Else, the partial sums of the Xbar are used as they are. 

\subsection{Hardware implementation of BinSparX}
\label{sec:hardware_implementation}

Fig~\ref{fig:binsparx}(b) provides an overview of a CiM-BNN memory array utilizing BinSparX.
There are four key hardware modifications to implement BinSparX. First, to support static weight sparsification, an additional register (1 bit per Xbar column) is required for the storage of $column\_flip$.   

Second, for dynamic activation sparsification, computing $\sum I$ requires an adder tree near-memory. Interestingly, the NAND-Net architecture already includes an adder tree for computing $\sum I$ for the post-processing, allowing us to utilize the existing hardware without significant hardware overheads. The additional circuitry needed includes a digital comparator (to compare $\sum I$ to  $\frac{n}{2}$) and XOR gates (1 per row) to flip $I$ or not, depending on $activation\_flip$. Additionally, when flipping $I$, the $\sum I$ required for post-processing (Equation~\ref{eq:nand_net}) becomes $n - \sum I$ due to the flip. To provide the correct value for post-processing, we incorporate a subtractor and a multiplexer to switch between $\sum I$ and $n - \sum I$ based on the value of $activation\_flip$.

Third, BinSparX needs some additional circuitry to post-process the partial sums. Recall, NAND-Net requires the VMM output to undergo some post-processing to obtain the final VMM output (equation~\ref{eq:nand_net}).  On top of that, BinSparX requires selective negation of partial sums based on the values of $activation\_flip$ and $column\_flip$. If $column\_flip[i]$ is 1 and $activation\_flip$ is 0 or vise versa, then the post-processed VMM output must be further multiplied by -1. However, if \textit{both} $activation\_flip$ and $column\_flip$ are 1 or 0, then no further processing is needed, since $\sum I*W = \sum (-I)*(-W)$.  Thus, the VMM output has to be multiplied with $(-1)^{(activation\_flip \oplus column\_flip)}$. We implement this using a XOR gate to calculate $activation\_flip \oplus column\_flip$, 2's complement circuitry to perform multiplication with -1, and a multiplexer to choose the final VMM output, with $activation\_flip \oplus column\_flip$ as the select signal.    

Fourth, BinSparX, interestingly, offers an opportunity to offset some of the overheads of the additional hardware discussed above. Typically for an Xbar where $n$ rows are asserted in parallel, $\log(n)$ bit analog to digital converters (ADCs) are required to digitize the analog current/voltage. However, static weight and dynamic activation sparsification ensure that Xbar weight columns and activations have $\leq \frac{n}{2}$ 1's, respectively. As a result, the VMM outputs are guaranteed to be $\leq \frac{n}{2}$. This allows the use of a $\log(n) - 1$ bit ADC for digitization without introducing any errors, leading to some energy, latency, and area savings. We will quantify the overall hardware benefits/overheads of BinSparX in Section~\ref{sec:hardware_overhead}. 

\begin{figure}[t!]
\centering
  \includegraphics[width = 0.95\linewidth]{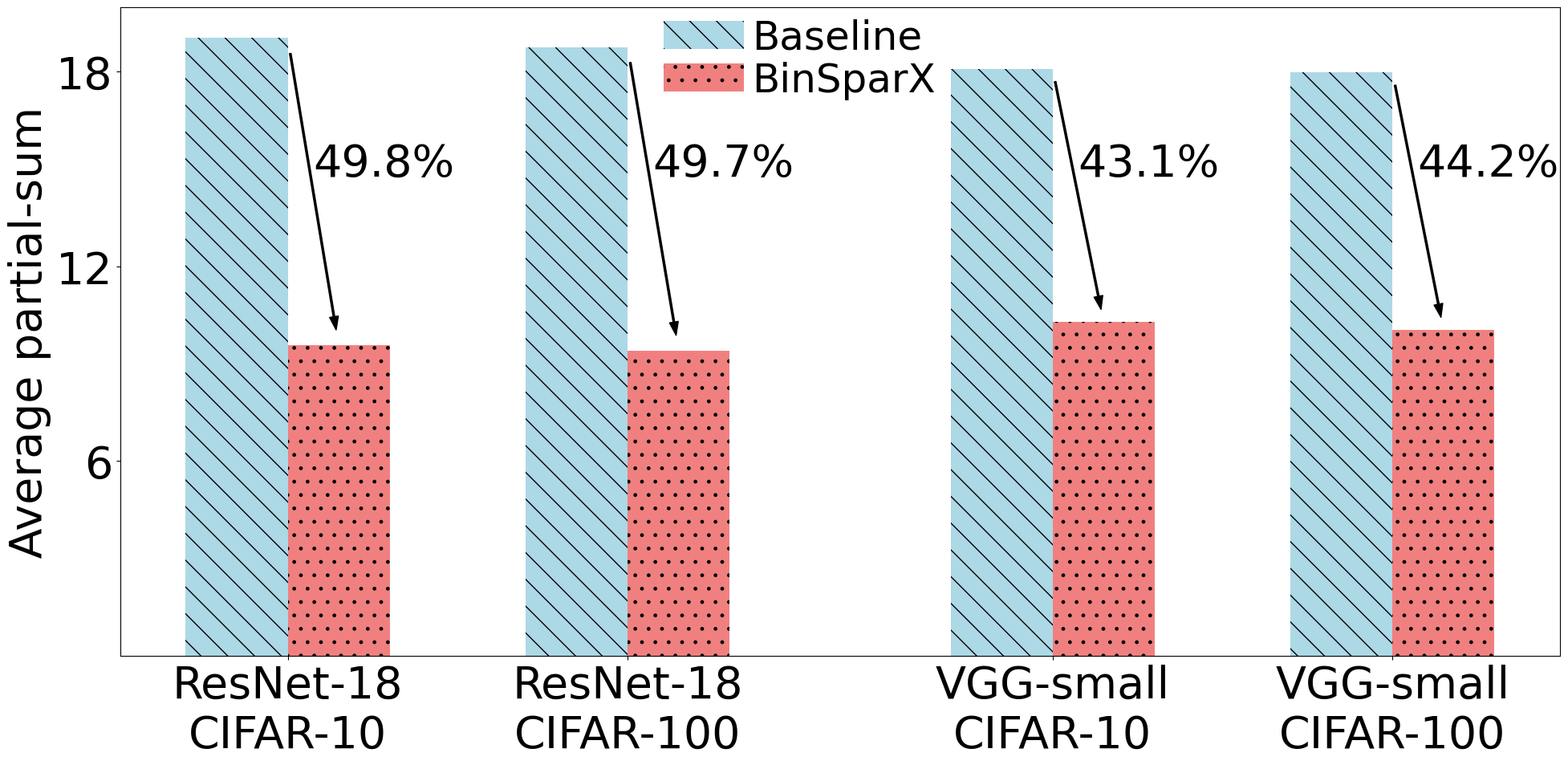}
  \caption{Average partial-sum reduction using BinSparX.}
  \label{fig:binsparx_reduction}
\end{figure}

\begin{figure*}[t!]
\centering
  \includegraphics[width = 0.98\textwidth]{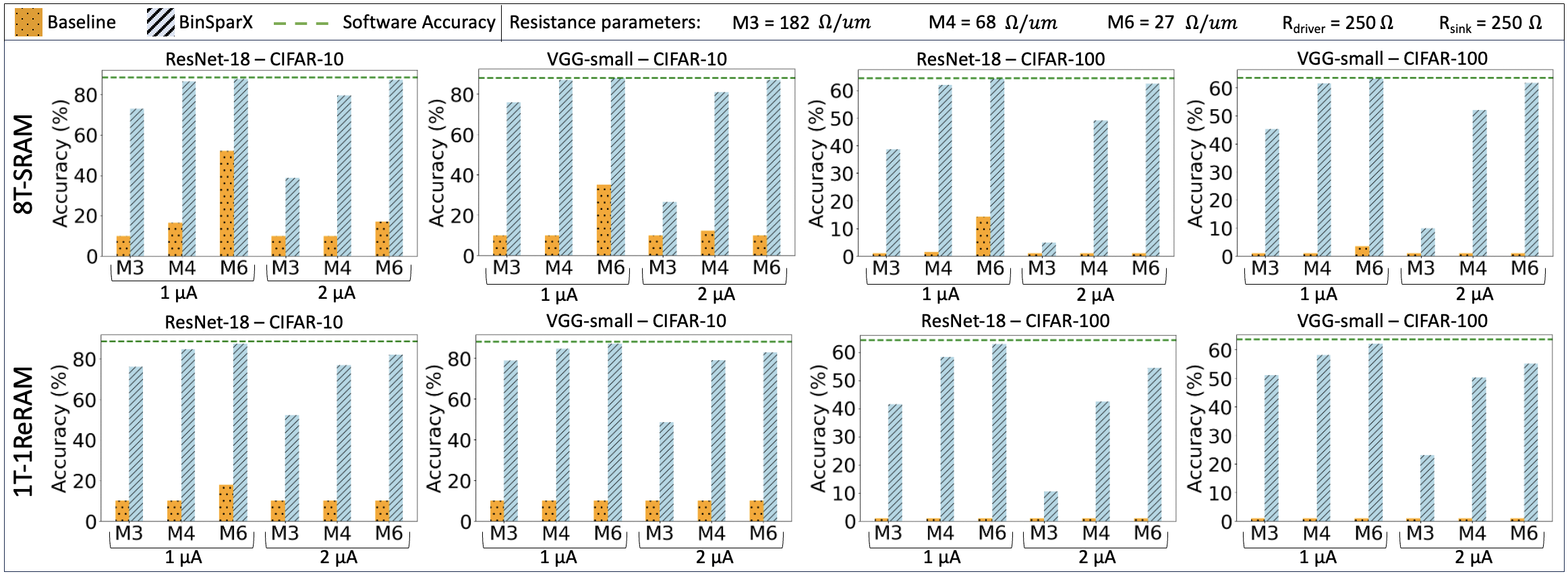}
  \captionsetup{justification=centering}
  \caption{Non-ideal inference accuracy for ResNet-18 and VGG-small CiM-BNNs across various hardware design points. Baseline CiM-BNN accuracy is severely degraded due to Xbar non-idealities, due to the large partial-sums. BinSparX is able to significantly improve accuracy, even bringing to near-ideal values for some cases. M3, M4 and M6 refer to the BL/SL routing metal layers, and 1 $\mu$A and 2 $\mu$A refer to the ON currents of the bit-cell. The 1T-ReRAM HRS current is $\sim$ 0.1 $\mu$A. We modulate the gap length in 1T-1ReRAM and the terminal voltages in 8T-SRAM to achieve the two different on-currents.}
  \label{fig:accuracy}
\end{figure*}
\section{Results}
\label{sec:results}

In this section, we evaluate the efficacy of BinSparX by testing it on ResNet-18 \cite{resnet18} and VGG-small \cite{vgg_small} BNNs, trained on CIFAR-10 and CIFAR-100. 

\subsection{Partial-sum reduction}
 First, we examine the partial-sum reduction achieved by BinSparX for the above-mentioned workloads. For this evaluation, we apply the methodology outlined in Section~\ref{sec:partial_sum_analysis}. Fig.~\ref{fig:binsparx_reduction} presents the average partial-sums for various workloads with and without BinSparX (baseline). Our results show that BinSparX reduces the average partial-sum magnitude by 43.1\% - 49.8\%. This sizeable reduction can be attributed to the multiplicative interaction between activations ($I$) and weights ($W$) in the Xbar. Since a bitcell is 'ON' only when both $I$ and $W$ are 1, transforming either $I$ or $W$ to 0 reduces the number of 'ON' bitcells. Consequently, the combined effects of static weight and dynamic activation sparsification lead to a large reduction in partial-sum magnitude. 

\subsection{Accuracy analysis}
Next, we examine the accuracy of BNNs when deployed on non-ideal Xbars, and quantify the accuracy improvements provided by BinSparX. We utilize the framework described in Section~\ref{sec:non_idealities_framework} for our analysis. We evaluate the accuracies for two bitcell 'ON' currents: 1 $\mu$A and 2 $\mu$A. The ON currents are controlled via $V_{DD}$ tuning in 8T-SRAM and filament tunneling gap tuning in ReRAMs.  While designs with higher 'ON current' offer larger distinguishability of output states and mitigate the effect of $I_{HRS/OFF}$ (especially in ReRAM), they also suffer from larger IR drops in the parasitic resistances.  Further, we assess the accuracies for BL/SL routed in M3, M4, and M6 metal layers. Higher metal layers, such as M6, have larger width and height, leading to lower wire resistance and reduced non-idealities, albeit at the cost of array area and capacitance (energy). We evaluate BinSparX across these design points to shows its efficacies for various design choices. 

Fig.\ref{fig:accuracy} summarizes the comparisons. The baseline accuracy (without BinSparX) is \textit{severely} degraded by Xbar non-idealities, with maximum accuracies of 52.09\% for CIFAR-10 and 18.08\% for CIFAR-100 (for favorable design points such as  M6 - 1 $\mu$A). For several other design points, near random-guess accuracies (10\% for CIFAR-10 and 1\% for CIFAR-100) are observed. 
On the other hand, BinSparX provides significant accuracy improvements. For the M6 - 1$\mu$A case, BinSparX helps attain accuracies within 0.5\% of the ideal accuracies for all CiM-BNNs. Even for  M6 - 2 $\mu$A and M4 - 1$\mu$A , the accuracies are increased to within 10\% of the ideal values. Thus, the partial-sum reduction (Fig.~\ref{fig:binsparx_reduction}) translates to significant accuracy gains. However, we observe that for the cases with larger non-idealities, e.g. M3 - 2 $\mu$A, the accuracy improvements may not be sufficient (but are still larger than the baseline). For such cases, BinSparX needs to be used in conjunction with with other non-ideality mitigation solutions.

\subsection{Hardware overhead}
\label{sec:hardware_overhead}

We evaluate the hardware implications of BinSparX at the memory macro level (Xbar + peripheral circuits). The actual impact may be lower once other components of the system are included. We evaluate the energy, latency and area of 64x64 Xbar arrays based on 8T-SRAM and 1T-1ReRAM using SPICE simulations. Note that the BL and SL of the Xbars are routed in M3 for the hardware evaluation. We utilize NeuroSim to estimate the energy-latency-area of the Xbar peripherals in the 7nm technology node \cite{neurosim} . For the baseline and BinSparX, we utilize 6-bit and 5-bit SAR-ADCs, respectively, based on our discussion in Section~\ref{sec:hardware_implementation}. We conduct our hardware evaluation for two cases: 1) each Xbar column is equipped with a dedicated ADC (64 ADCs), resulting in reduced latency but larger area and 2) 8 Xbar columns share a single ADC (8 ADCs), leading to a smaller area but higher latency.    

Table~\ref{table:overheads} summarizes the results. We observe that despite the additional hardware (See Section IV), BinSparX offers a 9-9.4\% \textit{reduction} in energy. This is because of the SAR-ADC bit precision reduction that offsets the effect of added peripherals. 

The latency is governed by two opposing factors. First, there is an improvement in latency due to the reduced ADC precision. However, in the baseline, the output of the adder tree is only needed for post-processing, allowing its latency to be masked by the CiM latency. In contrast, BinSparX needs the adder tree output for dynamic activation sparsification, adding the adder tree latency to the critical path. For the design with 64 ADCs per array, the adder tree latency dominates, leading to a 14.8–16.3\% increase in latency. Conversely, when 8 ADCs are used per array, the ADC latency becomes the bottleneck, resulting in a latency reduction of 8.8–9.7\%.

\begin{table}[t!]
\centering
\captionsetup{justification=centering}
\caption{Hardware overhead of BinSparX}
\footnotesize
\begin{tabular}{|p{1.5cm}|p{0.7cm}|c|c|c|c|}
\hline
\multirow{2}{*}{Memory} & \multirow{2}{*}{Energy} & \multicolumn{2}{c|}{Latency} & \multicolumn{2}{c|}{Area} \\
\cline{3-6}
 & & 8 ADCs & 64 ADCs & 8 ADCs & 64 ADCs \\
\hline
8T-SRAM & -9.0\% & -8.8\% & 14.8\% & 4.9\% & 3.4\% \\
\hline
1T-1ReRAM & -9.4\% & -9.7\% & 16.3\% & 6.0\% & 3.6\% \\
\hline
\end{tabular}
\label{table:overheads}
\end{table}

Lastly, we see that the area overhead of BinSparX, mainly from the extra register for storing the $column\_flip$ vector, is minimal, staying within 6\% across all evaluated design points. Thus, BinSparX offers significant accuracy improvements in CiM-BNNs with only minor hardware overhead.

\section{Conclusion}
\label{sec:conclusion}

In this work, we examined the impact of Xbar non-idealities on the accuracy of CiM-BNN accelerators designed in scaled technology nodes. We showed that inherently, BNNs produce larger partial sums compared to higher precision DNNs, due to which the influence of Xbar non-idealities in CiM-BNNs is more pronounced. To address this challenge, we proposed BinSparX—a training-free technique designed to mitigate the Xbar non-idealities specifically for CiM-BNNs. BinSparX reduces partial sums and, consequently, the non-idealities in BNN Xbar arrays by employing two key operations: static weight sparsification and dynamic activation sparsification. Our experiments conducted on ResNet-18 and VGG-small BNNs demonstrate that BinSparX achieves a 43.1\% to 49.8\% reduction in average partial-sum magnitudes. This leads to substantial accuracy improvements over a standard BNN, with accuracy gains reaching up to 77.25\%. Furthermore, we assessed the hardware implications of BinSparX and showed that the accuracy improvements are accompanied by a 9.0\% to 9.4\% reduction in energy consumption compared to the baseline. These benefits come at the cost of up to a 16.3\% and 6.0\% increase in latency and area, respectively.

\bibliography{references.bib}
\bibliographystyle{IEEEtran}

\end{document}